\documentclass[aps,pra,reprint,superscriptaddress]{revtex4-1} 

\usepackage{amsmath,amssymb} 
\usepackage{graphicx,xcolor} 
\usepackage{dcolumn} 
\usepackage{bm} 
\usepackage{bbm}
\usepackage{subfigure}
\usepackage{epstopdf}

\allowdisplaybreaks
\usepackage[utf8]{inputenc}
\usepackage[T1]{fontenc}

\begin{document}

\title{Improved low-dimensional wave equations for cigar-shaped and disk-shaped dipolar Bose-Einstein condensates}
\author{Mitchell J. Knight}
\affiliation{The University of Melbourne, School of Physics, Parkville, Victoria 3010, Australia}

\author{Thomas Bland}
\affiliation{Joint Quantum Centre (JQC) Durham-Newcastle, School of Mathematics, Statistics and Physics, Newcastle University, Newcastle upon Tyne, NE1 7RU, United Kingdom}

\author{Nick G. Parker}
\affiliation{Joint Quantum Centre (JQC) Durham-Newcastle, School of Mathematics, Statistics and Physics, Newcastle University, Newcastle upon Tyne, NE1 7RU, United Kingdom}

\author{Andy M. Martin}
\affiliation{The University of Melbourne, School of Physics, Parkville, Victoria 3010, Australia}

\date{\today}

\begin{abstract}
Within the formalism of the Gross-Pitaevskii equation, we derive effective one- and two-dimensional equations for cigar- and pancake-shaped dipolar Bose-Einstein condensates with arbitrary polarization angle.  These are based on an  ansatz for the condensate wavefunction whose width in the tightly-confined direction/s is treated variationally.  The equations constitute a coupled partial differential for the low-dimensional wavefunction and algebraic equations for the width parameters.  This approach accurately predicts the ground state densities of cigar-shaped and pancake-shaped dipolar Bose-Einstein condensates, and gives strong agreement with the three-dimensional results, even as the trapping is relaxed away from the strict quasi-one- and quasi-two-dimensional regimes.  This approach offers a significant improvement over the standard one- and two-dimensional reduction.
\end{abstract}

\maketitle

\section{Introduction}

Gaseous Bose-Einstein condensates (BECs) are today routinely created by 
 experimental groups 
 over the world. In the decade following the creation of the first BEC in 1995 \cite{Science269.198}, typical BECs were made from 
 alkali metals with small magnetic dipole moments, such as lithium-7 \cite{PhysRevLett.78.985}, sodium-23 \cite{PhysRevLett.75.3969}, and potassium-41 \cite{Science294.1320}. In such BECs, the atom-atom interactions are short-range and isotropic, arising from van der Waals forces, and are well accounted for theoretically as a contact pseudo-potential \cite{StringariPitaevskii}. In more recent times, BECs have also been produced using atoms with significant magnetic dipole moments. This began with the condensation of chromium-52 in 2005 \cite{PhysRevLett.94.160401}, followed by dysprosium-164 \cite{PhysRevLett.107.190401}, erbium-168 \cite{PhysRevLett.108.210401}, and erbium-166 \cite{PhysRevX.6.041039}.  In these so-called dipolar BECs, the magnetic dipoles are polarized in a common direction by an externally-applied magnetic field.  Alongside this experimental effort there has been a renewed interest in the theoretical understanding of 
 dipolar BECs, as well as dipolar quantum gases more generally
\cite{Lahaye_2009,Baranov2012}.

In contrast to the van der Waals/contact interactions, the dipole-dipole interaction (DDI) between the atoms is anisotropic (part attractive and part repulsive depending on the relative orientation of the dipoles) and long-range.
These characteristics underpin a range of striking physical phenomena, such as magnetostriction \cite{PhysRevA.63.053607,PhysRevLett.85.1791,PhysRevA.61.051601,PhysRevA.73.031602,Lahaye2007}, the existence of a roton dip in the excitation spectrum \cite{PhysRevLett.90.110402,PhysRevLett.90.250403,Chomaz}, and roton-driven density corrugations \cite{PhysRevA.75.053604,PhysRevLett.100.245302,PhysRevA.82.023622} and recently-reported supersolid phases \cite{PhysRevLett.122.130405,PhysRevX.9.011051,PhysRevX.9.021012}.  

\begin{figure}[h]
	\includegraphics[width=0.7\linewidth]{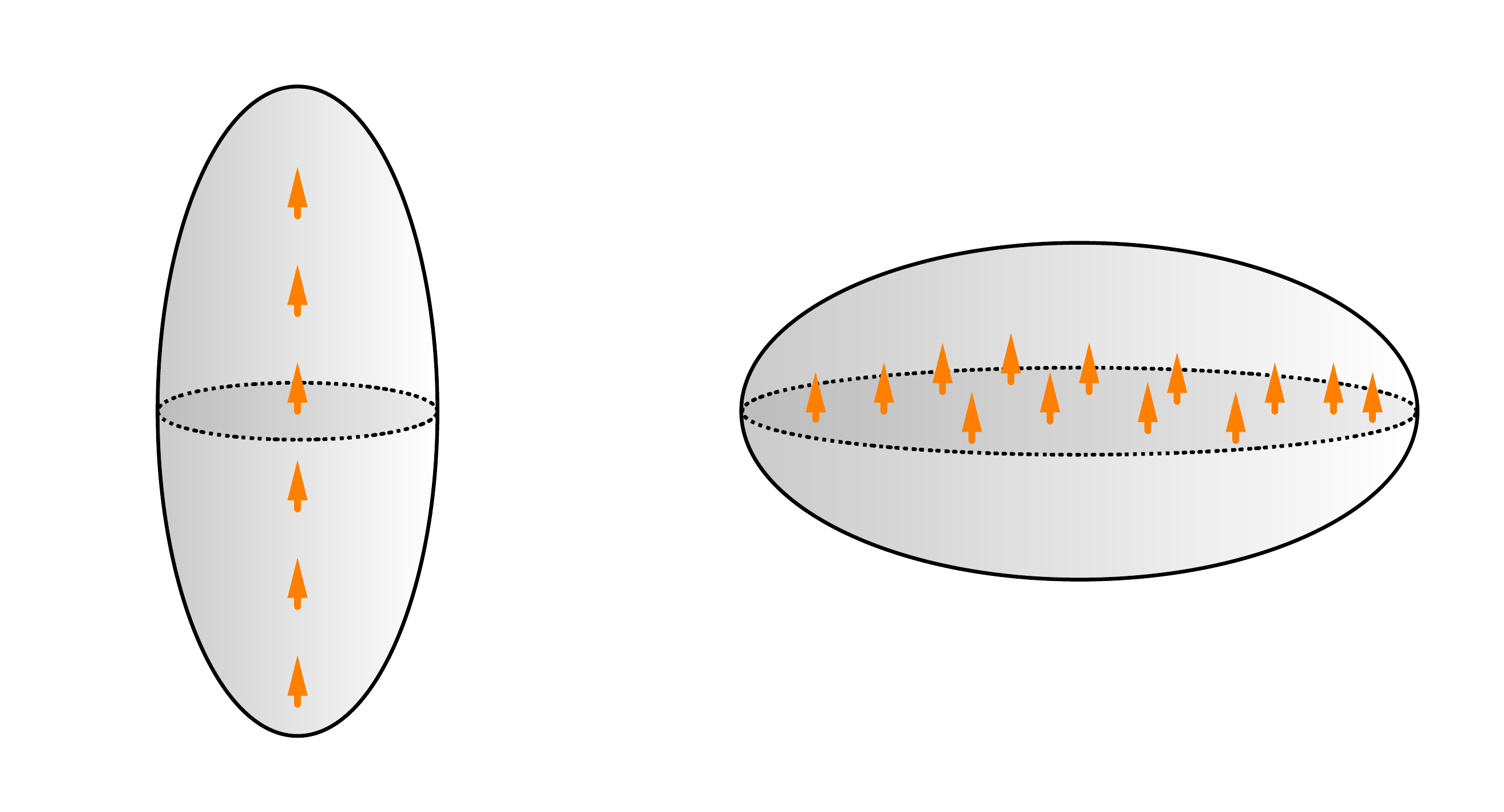}
	\caption{The geometry of the trapping potential can augment or reduce the magnitude of the DDIs. In the cigar-shaped trap (left-hand-side) the attractive head-to-tail configuration is favoured, whereas in the pancake-shaped trap (right-hand-side) the repulsive side-to-side configuration is favoured. }
	\label{fig:diagram1}
\end{figure}

There is strong interest in dipolar BECs in elongated and flattened geometries, for a variety of reasons.  Firstly, such geometries can be exploited to enhance the role of the DDIs, as shown in Fig.~\ref{fig:diagram1}: in an elongated dipolar BEC with dipoles polarised along the primary axis, the attractive head-to-tail alignment is favoured, whereas in a flattened geometry, the repulsive side-by-side configuration is favoured.  Secondly, strong trapping in one or two dimensions is required to support the presence of the roton and its rich manifestations mentioned above.  Thirdly, such geometries allow access to dipolar physics in lower dimensions, including one-dimensional bright solitons \cite{PhysRevA.79.053608, PhysRevA.92.033605, Edmonds_2017}, two-dimensional bright solitons \cite{PhysRevLett.95.200404, PhysRevA.93.023633,PhysRevA.96.053617,PhysRevLett.100.090406,PhysRevA.92.013637,PhysRevA.93.053608}, one-dimensional dark solitons \cite{Pawlowski_2015, PhysRevA.92.063601, PhysRevA.95.063622, PhysRevA.93.063617, Umarov_2016}, vortices and vortex lattices \cite{PhysRevLett.95.200402, PhysRevLett.95.200403, PhysRevA.73.061602, PhysRevA.79.013621, PhysRevA.79.063622, PhysRevA.83.033628, PhysRevLett.111.170402, Gautam_2014, Martin_2017, PhysRevA.96.063624, PhysRevA.97.043614, PhysRevA.98.023610, Prasad2019}, and near-integrability \cite{PhysRevX.8.021030}.

The workhorse model for BECs in general is the Gross-Pitaevskii equation (GPE), a zero-temperature mean-field model for the macroscopic wavefunction of the BEC \cite{Pitaevskii,Gross1961,StringariPitaevskii,pethick_smith_2008,Barenghi}.  In the absence of DDIs, it has the form of a cubic nonlinear Schr\"odinger equation (NLSE), while it can be extended to dipolar BECs through the inclusion of an additional, non-local term \cite{Lahaye_2009}.  Theoretical studies of low dimensional BECs typically employ the reduced 1D or 2D forms of the GPE, in which the condensate wavefunction is written as the product of the low dimensional wavefunction and a static profile (the non-interacting harmonic oscillator state) in the tightly confined direction(s).  By integrating out the latter, one arrives at the effective 1D/2D GPE.  These equations are strictly valid in the so-called quasi-1D and quasi-2D regimes.  Assuming the condensate to by cylindrically symmetric, with radius $R_\perp$ and axial half-length $R_z$, then the quasi-1D regime requires $R_z > \xi > R_\perp$, where $\xi$ is the healing length of the condensate.  Similarly, the quasi-2D regime requires $R_\perp > \xi > R_z$.  The form and validity of these equations have been established for conventional BECs \cite{PhysRevLett.81.938,PhysRevLett.84.2551,PhysRevLett.85.3745,PhysRevA.66.043610,PhysRevA.74.065602,StringariPitaevskii,Barenghi}, and later extended to dipolar BECs \cite{PhysRevA.78.041601,PhysRevA.82.043623}.

The widespread use of the reduced 1D and 2D GPEs is motivated by their simplicity and ease of solution.  However, their predictive power in capturing realistic 3D behaviour has its limits: the strict quasi-1D and quasi-2D regimes in which they are valid can be challenging to reach experimentally and they are not quantitatively accurate in more relaxed geometries.  Moreover, 1D and 2D models can fail to capture important physical effects, with the notable example being the collapse instability of attractively-interacting BECs, which is not captured by the 1D GPE \cite{PhysRevA.56.1424,PhysRevA.66.043603,Parker_2007}.  


For conventional (non-dipolar) BECs, low dimensional equations have been pursued which capture some of the residual three-dimensional physics and overcome the above challenges \cite{PhysRevA.65.043614}.  These so-called non-polynomial Schr\"odinger equations take the profile of the wavefunction in the tightly-confined direction(s) as a Gaussian ansatz with variable width, rather than fixed width.  They provide excellent agreement with the 3D solutions in cigar-shaped and pancake-shaped condensates, even if the tight confinement is relaxed away from the strict quasi-1D and quasi-2D regime.  Moreover, the 1D non-polynomial GPE successfully captured the 3D physics of the collapse instability \cite{PhysRevA.66.043603}.  

In the dipolar case, the strict 1D/2D GPEs have been derived and shown to capture the 3D behaviour when the system is deep in the 1D/2D regime, but depart away from this regime \cite{PhysRevA.82.043623,Baillie_2015}.  To date, the above approach has not been extended to the dipolar BECs.  One can expect that such equations would allow for accurate, convenient prediction of elongated and flattened dipolar BECs over a greater parameter regime, and may be particularly important in capturing the three-dimensional physics of the collapse and rotons.

Motivated by this, in this work we use the variable-width approach to derive effective lower-dimensional equations for cigar-shaped and pancake-shaped dipolar BECs, in the manner of Salasnich {\it et al.~}\cite{PhysRevA.82.043623}. Because both the lower-dimensional wavefunction and the width are variational, we minimise the energy of the system with respect to both, and arrive at a wave equation for the wavefunction, and a polynomial equation for the width/s. Solving these coupled equations in tandem allows for an accurate prediction of the system.  We will show that these equations accurately predict the ground state solutions, and provide a significant improvement over the standard 1D/2D effective equations.  In Section II we describe the over-arching theoretical model of the 3D dipolar GPE, while in Sections III and IV we proceed to derive and test the variable-width 1D and 2D equations, comparing to solutions of the full 3D dipolar GPE.  In Section V we conclude our work.

\section{The 3D Model}

\begin{figure}[h]
	\centering
	\raisebox{0.5\height}{\subfigure[]{\includegraphics[scale=0.25]{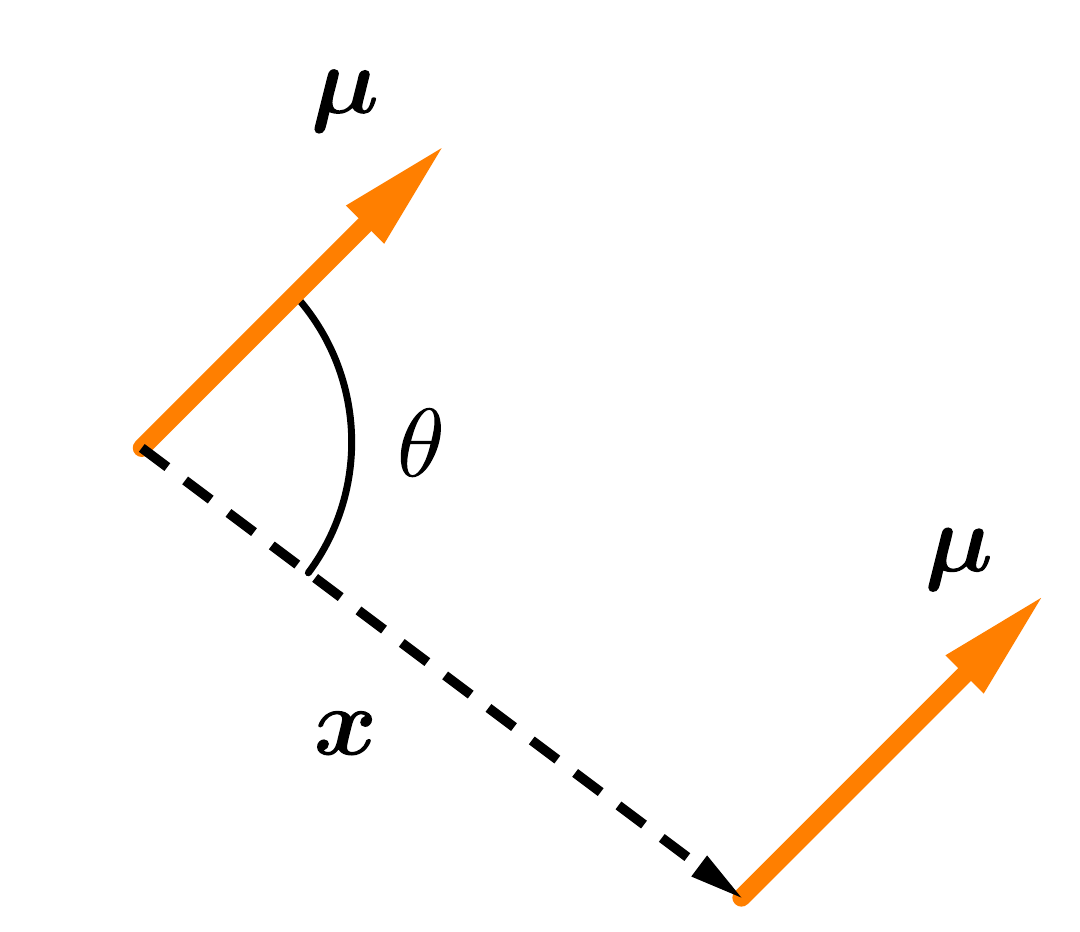}}}\quad
	\subfigure[]{\includegraphics[scale=0.25]{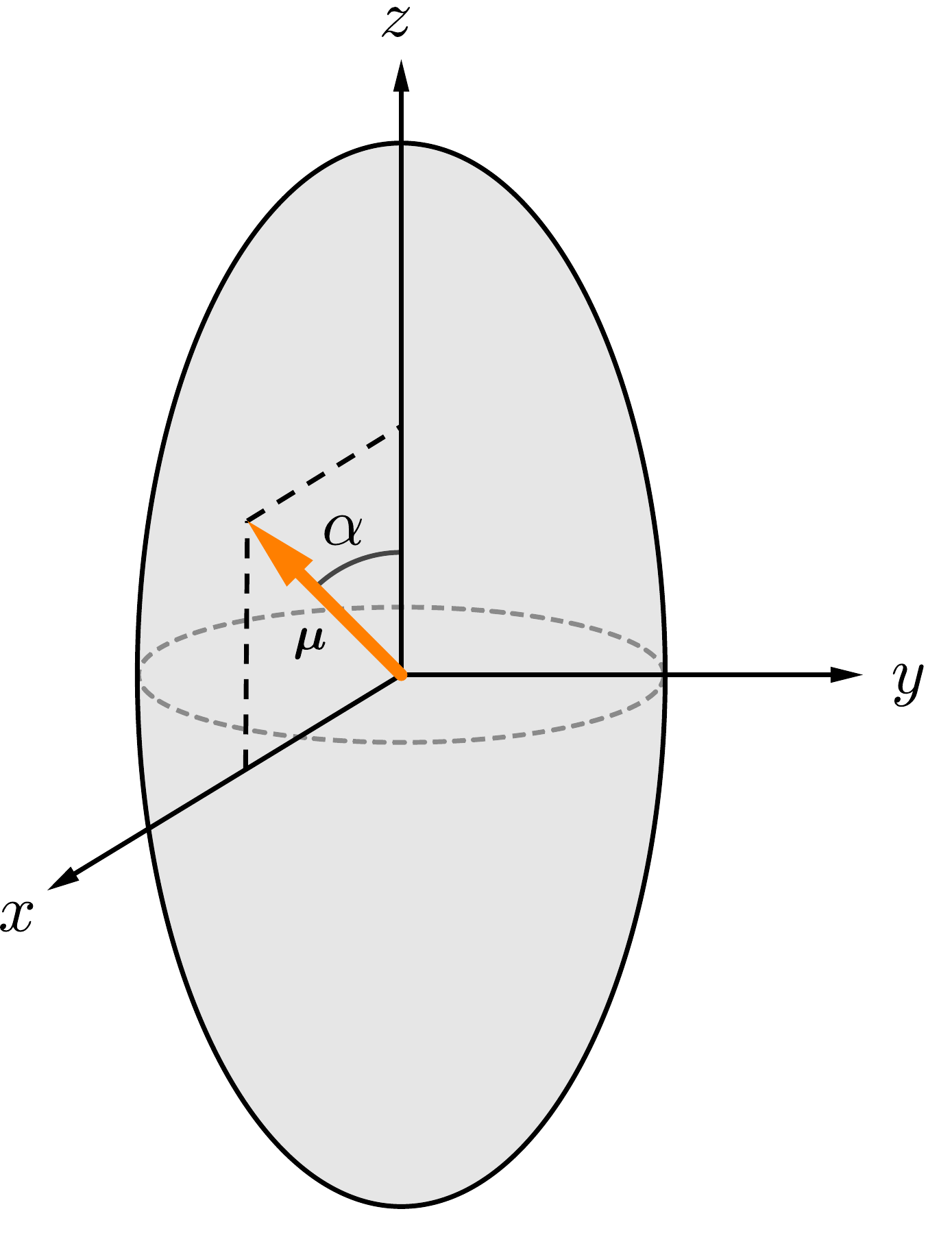}}
	\caption{(a)  The polarisation vector $\bm \mu$ and the inter-particle vector $\bm x$, with angle $\theta$ between them. (b) The polarisation angle $\alpha$ as measured from the positive $z$-axis in the $xz$-plane.}
	\label{fig:diagram23}
\end{figure}

Consider a dipolar BEC of $N$ atoms at zero temperature, confined in an external trapping potential $\mathcal{V}_{\text{trap}}(\bm x)$. The time-evolution of the macroscopic wavefunction, $\psi(\bm x,t)$, is given by the 3D dipolar GPE (dGPE) \cite{Lahaye_2009}
\begin{multline}
i\hbar\partial_t\psi(\bm x,t)=\Bigg\{-\frac{\hbar^2}{2m}\nabla^2+\mathcal{V}_{\text{trap}}(\bm x)+gN|\psi(\bm x,t)|^2 \\ +\Phi_{\text{dd}}(\bm x)\Bigg\}\psi(\bm x,t), \label{eq:3dgpe}
\end{multline}
where $\nabla^2$ is the Laplacian, $g=4\pi\hbar^2a_s/m$ is the $s$-wave coupling constant with $a_s$ being the $s$-wave scattering length and $m$ being the mass of the constituent bosons. $|\psi(\bm x,t)|^2$ is the probability density of the atoms, satisfying the normalisation condition $\int\text{d}\bm x\,|\psi(\bm x,t)|^2=1$. $\Phi_{\text{dd}}$ is the mean-field potential generated by the DDIs, which is given by the convolution
\begin{equation}
\Phi_{\text{dd}}(\bm x)=N\int\text{d}\bm x'\,\mathcal{V}_{\text{dd}}(\bm x-\bm x')|\psi(\bm x',t)|^2, \label{eq:ddipotential}
\end{equation}
where $\mathcal{V}_{\text{dd}}(\bm x)$ is the DDI interaction term
\begin{equation}
\mathcal{V}_{\text{dd}}(\bm x)=\frac{C_{\text{dd}}}{4\pi}\frac{1-3\cos^2\theta}{|\bm x|^3}. \label{eq:ddiinteraction}
\end{equation}
Here $\theta$ is the angle between the polarisation direction of the dipoles and the inter-particle vector between two interacting dipoles (see Fig.~\ref{fig:diagram23}) and $C_{\text{dd}}=\mu_0\mu_{\text{d}}^2$, where $\mu_0$ is the magnetic vacuum permeability and $\mu_{\text{d}}$ is the magnetic dipole moment of the bosons. We introduce the dimensionless parameter $\epsilon_{\text{dd}}=C_{\text{dd}}/(3g)$ which parametrises the strength of the DDIs relative to the contact interactions. 

The 3D dGPE can be recast into a more useful form as the Euler-Lagrange equation of the action functional of the system. Such an approach results in a clear method of dimension reduction, as shown by Salasnich {\it et al.~}\cite{PhysRevA.65.043614}. The action functional of the system is 
\begin{multline}
\mathcal{S}=\int\text{d}t\int\text{d}\bm x\,\psi^*(\bm x,t)\Bigg\{i\hbar\partial_t+\frac{\hbar^2}{2m}\nabla^2-\mathcal{V}_{\text{trap}}(\bm x) \\ -\frac{1}{2}gN|\psi(\bm x,t)|^2-\frac{1}{2}\Phi_{\text{dd}}(\bm x)\Bigg\}\psi(\bm x,t), \label{eq:3dactionfunctional}
\end{multline}
where $\psi^*(\bm x,t)$ is the complex conjugate of $\psi(\bm x,t)$. The 3D dGPE in Eq.~\eqref{eq:3dgpe} is the Euler-Lagrange equation of Eq.~\eqref{eq:3dactionfunctional}, taken with respect to $\psi^*(\bm x,t)$. We recast the mean-field dipolar potential through the convolution theorem
\begin{equation}
\Phi_{\text{dd}}(\bm x)=\mathcal{F}^{-1}\Big\{\tilde{\mathcal{V}}_{\text{dd}}(\bm k)\,\mathcal{F}[|\psi(\bm x,t)|^2]\Big\}, \label{eq:ddipotentialFourier}
\end{equation}
where $\mathcal{F}$ ($\mathcal{F}^{-1}$) denotes the Fourier (inverse Fourier) transform, and $\tilde{\mathcal{V}}_{\text{dd}}(\bm k)$ is the Fourier transform of the DDI, given by \cite{Lahaye_2009} 
\begin{equation}
\frac{\tilde{\mathcal{V}}(\bm k)}{C_{\text{dd}}}=\frac{k_x^2\sin^2\alpha+k_xk_z\sin(2\alpha)+k_z^2\cos^2\alpha}{k_x^2+k_y^2+k_z^2}-\frac{1}{3},\label{eq:ddiinteractionFourier}
\end{equation}
where $\alpha$ is the angle of polarisation of the dipoles measured from the positive $z$-axis in the $xz$-plane (see Fig.~\ref{fig:diagram23}).

The dipolar BEC is confined in a cylindrically symmetric harmonic oscillator trap with trap frequency $\omega_z$ in the $z$ direction, and $\omega_{\perp}$ in the $x$- and $y$-directions,
\begin{equation}
\mathcal{V}_{\text{trap}}(\bm x)=\frac{1}{2}m\omega_{\perp}^2(x^2+y^2)+\frac{1}{2}m\omega_z^2z^2, \label{eq:trap}
\end{equation}
For convenience we define the trap ratio $\gamma=\omega_{\perp}/\omega_z$: a cigar-shaped trap corresponds to $\gamma\gg 1$ and a pancake-shaped trap to $0<\gamma\ll1$. We also introduce the harmonic oscillator lengths, $\ell_z=\sqrt{\hbar/m \omega_z}$ and $\ell_\perp=\sqrt{\hbar/m\omega_\perp}$.

The accuracy of our low-dimensional equations is found by comparing their ground state solution to the 3D result.  The 3D ground state solutions are found via imaginary-time propagation of the 3D dGPE using the split-step Fourier method \cite{fleck_morris_1976,fleck_morris_1978}.  We employ a 256$^3$ grid with discretization $(\Delta x,\Delta y,\Delta z)=(0.04,0.04,0.0008)\ell_\perp$ for pancake geometries, and $(\Delta x,\Delta y,\Delta z)=(0.008,0.008,0.04)\ell_\perp$ in cigar geometries. Since fast Fourier transform algorithms are naturally periodic, we introduce a cylindrical cut-off to the dipolar potential, which restricts the range of the DDI to radius $\rho=\sqrt{x^2+y^2}<\rho_c$ and length $|z|<z_c$ in real-space in order to reduce the effects of alias copies, giving a real-space DDI
\begin{align}
U_\text{dd}^{\rho_c,Z_c}(\bm r)=
\begin{cases}
\displaystyle\frac{C_\text{dd}}{4\pi}\frac{1-3\cos^2\theta}{r^3}\,,& \rho<\rho_c\text{ and }|z|<Z_c\,, \\
0\,, & \text{otherwise}.
\end{cases}
\end{align}
The Fourier transform of this is semi-analytic \cite{lu_lu_2010}, given by
\begin{align}
&\tilde{U}_\text{dd}^{\rho_c,Z_c}(\bm k)=\frac{C_\text{dd}}{3}\left(3\cos^2\theta_k-1\right)+C_\text{dd}e^{-Z_ck_\rho}F(\theta_k,z_k) \nonumber \\
&-C_\text{dd}\int_{\rho_c}^\infty\text{d}\rho\, \rho  \int_0^{Z_c}\text{d} z\, \cos(k_z z)\frac{\rho^2-2z^2}{\left(\rho^2+z^2\right)^{5/2}}J_0(k_\rho \rho)\,,
\end{align}
where $J_0$ is the zeroth-order Bessel function of the first kind and $F(\theta_k,z_k)=\sin^2\theta_k\cos\left(Z_ck_z\right)$ $-\sin\theta_k\cos\theta_k\sin\left(Z_ck_z\right)$. The final line requires numerical integration, which is computationally costly but only required once as part of the simulation setup.

\section{Variable-Width One-Dimensional Equations}

We assume a prolate harmonic trap $\gamma \gg 1$ giving rise to a cigar-shaped dipolar BEC.  Following Salasnich {\it et al.} \cite{PhysRevA.65.043614}, we write the 3D wavefunction in the separable form,
\begin{align}
\psi(\bm x,t)&=f(z,t)\varphi(x,y,t;\sigma(z,t)) 
\label{eq:1dwavefunction}
\end{align}
where $f$ is the 1D wavefunction along $z$ and $\varphi$ is taken to have a Gaussian profile in $x$ and $y$, whose width $\sigma$ varies in time and with $z$,  
\begin{align}
\varphi(x,y,t;\sigma(z,t))=\frac{\exp\left[\displaystyle\frac{-(x^2+y^2)}{2\sigma(z,t)^2}\right]}{\sqrt{\pi}\sigma(z,t)}
\end{align}
Note that both $f$ and $\varphi$ are normalised to unity. The strict 1D limit derived previously \cite{PhysRevA.82.043623} corresponds to the fixed width $\sigma=\ell_\perp$.  However, here we consider variable width, treating $\sigma$ as a variational parameter.


To derive the variable-width 1D equations, we substitute the above form of the wavefunction into the action functional of Eq.~\eqref{eq:3dactionfunctional}, along with the DDI potential \eqref{eq:ddipotentialFourier} and trapping potential \eqref{eq:trap}, and integrate out the $x$ and $y$ dimensions. 
We also assume that $\varphi$ is slowly varying along the $z$-direction, allowing us to write $\nabla^2\varphi\approx \nabla_{\perp}^2\varphi$ where $\nabla_{\perp}^2=\partial_x^2+\partial_y^2$.  We then arrive at the 1D action functional,
\begin{multline}
\mathcal{S}_{1\text{D}}=\int\text{d}z\int\text{d}t\,f^*\Bigg\{i\hbar\partial_t+\frac{\hbar^2}{2m}\partial_z^2-\frac{1}{2}m\omega_z^2z^2-\frac{\hbar^2}{2m\sigma^2} \\ -\frac{1}{2}m\omega_{\perp}^2\sigma^2-\frac{gN}{4\pi}\frac{|f|^2}{\sigma^2}-\frac{1}{2}\Phi_{\text{dd}}^{1\text{D}}\Bigg\}f.\label{eq:1dactionfunctional}
\end{multline}
Here, $\Phi_{\text{dd}}^{1\text{D}}$ is the 1D DDI potential which is derived from Eq.~\eqref{eq:ddiinteractionFourier} by evaluating the Fourier (inverse Fourier) transform in the $x$ and $y$ ($k_x$ and $k_y$) directions, the result of which depends upon the polarisation direction $\alpha$. $\Phi_{\text{dd}}^{1\text{D}}$ is only analytic for $\alpha=0$; for $\alpha\neq 0$, the evaluation of the Fourier and inverse Fourier transforms have to be completed numerically. 

\subsection{$z$-Polarisation Case}

If $\alpha=0$, the dipoles are aligned along the positive $z$-axis and share an axis with the primary axis of the cigar-shaped BEC. The 1D DDI potential then becomes,
\begin{equation}
\Phi_{\text{dd}}^{1\text{D}}\Bigg|_{\alpha=0}=\frac{C_{\text{dd}}N}{6\pi\sigma^2}\mathcal{F}_z^{-1}\Big\{(3qe^q\Gamma[0,q]-1)\mathcal{F}_z\{|f|^2\}\Big\}, \label{eq:1dddipotentialalpha0}
\end{equation}
where $q=2\pi^2k_z^2\sigma^2$, $\mathcal{F}_z$ ($\mathcal{F}_z^{-1}$) denotes the 1D Fourier (inverse Fourier) transform and $\Gamma[0,\cdot]$ is the incomplete gamma function. With this DDI potential in the 1D action functional in Eq.~\eqref{eq:1dactionfunctional}, we apply the Euler-Lagrange equations to the conjugate of the wavefunction, $f^*$, and the variational width, $\sigma$. This leads to the following system of equations, an integro-differential equation for $f(z)$ coupled to an equation for $\sigma$,
\begin{align}
i\hbar\partial_tf&=\Bigg\{-\frac{\hbar^2}{2m}\partial_z^2+\frac{1}{2}m\omega_z^2z^2+\frac{\hbar^2}{2m\sigma^2}+\frac{1}{2}m\omega_{\perp}^2\sigma^2\notag\\
& \hspace{80px}+\frac{gN|f|^2}{2\pi\sigma^2}+\Phi_{\text{dd}}^{1\text{D}}\Big|_{\alpha=0}\Bigg\}f,\label{eq:1ddiff1}\\
0&=\frac{\hbar^2}{2m\sigma^3}-m\omega_{\perp}^2\sigma+\frac{gN|f|^2}{2\pi\sigma^3}-\frac{\partial}{\partial\sigma}\Phi_{\text{dd}}^{1\text{D}}\Big|_{\alpha=0}.\label{eq:1dalg1}
\end{align}
These are the variable-width GPE equations for the cigar-shaped dipolar BEC with $\alpha=0$.

For convenience, we scale our variables into dimensionless form using harmonic oscillator units based on the trapping along $z$: the timescale is $\omega_z^{-1}$ and the lengthscale is $\ell_z$. The dimensionless variables (denoted by tilde) are then,
\begin{equation}
\tilde{t}=t\omega_z, \ \tilde{z}=\frac{z}{\ell_z}, \ \tilde{a}_s=\frac{a}{\ell_z}, \ \tilde{\sigma}=\frac{\sigma}{\ell_z},\ \tilde{k}_z=k_z\ell_z. \label{eq:1dunits}
\end{equation}
It follows that the dimensionless wavefunction is $\tilde{f}=f\sqrt{\ell_z}$ and the harmonic oscillator length in the $xy$ plane is $\tilde{\ell}_{\perp}=1/\sqrt{\gamma}$. With these transformations, the variable-width 1D equations for $\alpha=0$ become,
\begin{widetext}
\begin{align}
i\partial_{\tilde{t}}\tilde{f}&=\left\{-\frac{1}{2}\partial_{\tilde{z}^2}+\frac{1}{2\tilde{\sigma}^2}+\frac{\tilde{z}^2}{2}+\frac{\gamma^2\tilde{\sigma}^2}{2}+\frac{\tilde{\beta}_{1\text{D}}|\tilde{f}|^2}{\tilde{\sigma}^2}(1-\epsilon_{\text{dd}})+\frac{\epsilon_{\text{dd}}}{2}\mathcal{F}_{\tilde{z}}^{-1}\left((3qe^{q}\Gamma[0,q])\mathcal{F}_{\tilde{z}}\{|\tilde{f}|^2\}\right)\right\}\tilde{f}, \label{eq:1ddiff2}\\
0&=1-\gamma^2\tilde{\sigma}^2+\tilde{\beta}_{1\text{D}}\left[|\tilde{f}|^2-\epsilon_{\text{dd}}\mathcal{F}_{\tilde{z}}^{-1}\left((1-3q+3q^2e^{q}\Gamma[0,q])\mathcal{F}_{\tilde{z}}\{|\tilde{f}|^2\}\right)\right], \label{eq:1dalg2}
\end{align}
\end{widetext}
where $\tilde{\beta}_{1\text{D}}=2\tilde{a}_sN$ and $\mathcal{F}_{\tilde{z}}$ ($\mathcal{F}_{\tilde{z}}^{-1}$) is the scaled Fourier (inverse Fourier) transform. These equations have a similar form to the 3D dGPE given in Eq.~\eqref{eq:3dgpe}. The first and second terms in Eq.~\eqref{eq:1ddiff2} are the kinetic energy, and the third and fourth terms are the potential energy.  The final two terms are due to the interactions: the first of these is a contact interaction term combining the van der Waals interactions and the short-range component of the DDIs, while the second is the long-range interactions of the DDIs. 

For $\epsilon_{\text{dd}}=0$, Eq.~\eqref{eq:1dalg2} can be solved analytically for $\tilde{\sigma}$, and substitution into Eq.~\eqref{eq:1ddiff2} results in the NPSE of Salasnich {\it et al.~}\cite{PhysRevA.65.043614}. Such a simplification cannot be implemented for $\epsilon_{\text{dd}}\neq 0$, and therefore Eq.~\eqref{eq:1ddiff2} and Eq.~\eqref{eq:1dalg2} must be solved numerically in a self-consistent manner.

\begin{figure}[th!]
	\includegraphics[width=\linewidth]{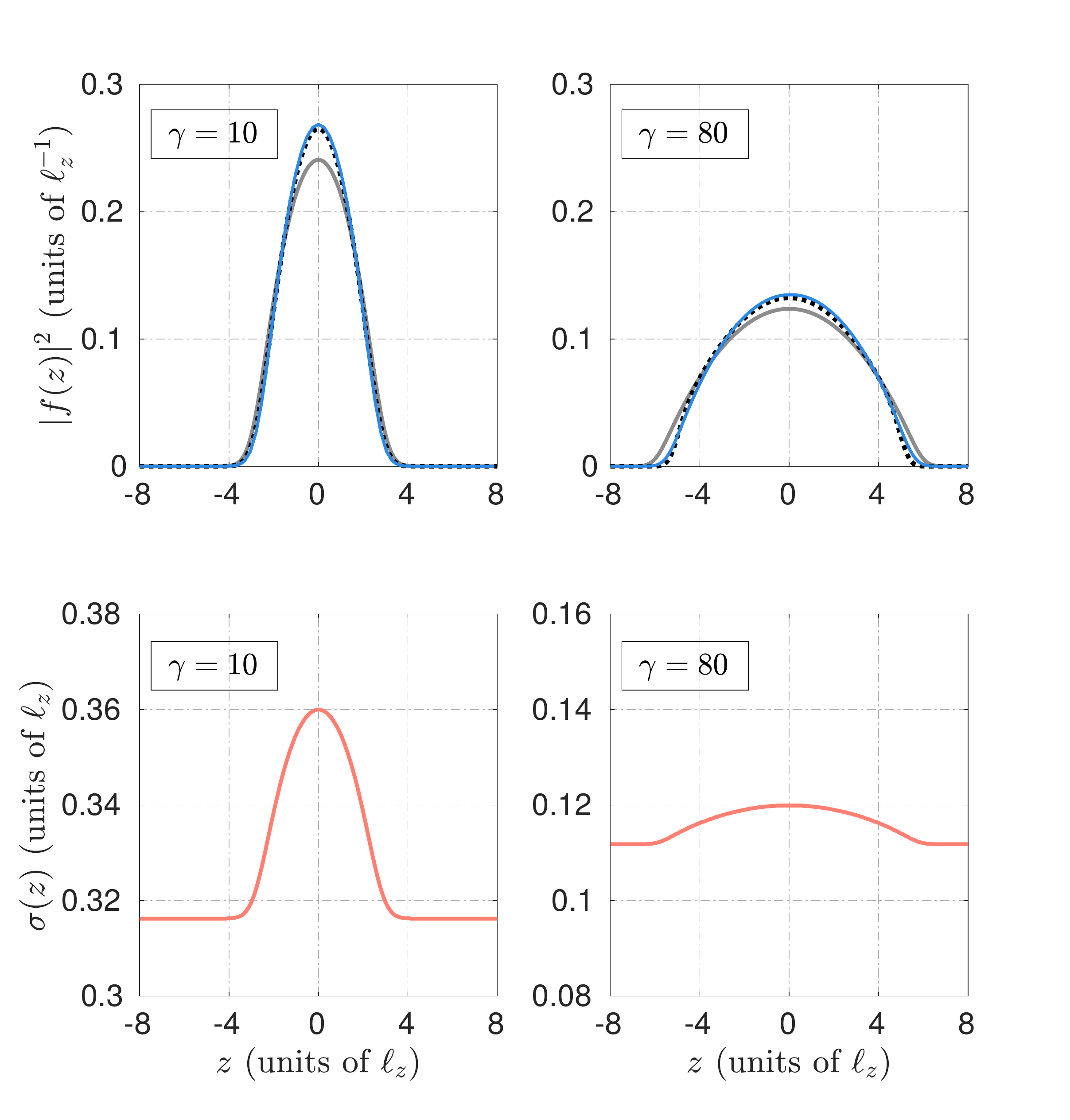}
	\caption{Cigar-shaped dipolar BECs with dipoles polarised along $z$.  Top panels:  Ground state density $|f(z)|^2$ according to the variable-width approach (dotted black line), the fixed-width approach (solid grey line) and the 3D dGPE (solid blue line). Bottom panels:  Variational width $\sigma(z)$ according to Eq.~\eqref{eq:1dalg2}. We take $\epsilon_{\text{dd}}=0.9$ and $\tilde{\beta}_{1\text{D}}=2$.}
	\label{fig:result1}
\end{figure}

In the top panels of Fig.~\ref{fig:result1} we plot the ground state density $|f(z)|^2$ as predicted by the variable-width 1D model [Eqs.~(\ref{eq:1ddiff2}) and (\ref{eq:1dalg2})], the fixed-width model [Eq.~\eqref{eq:1ddiff2} with $\tilde{\sigma}=\ell_\perp$], and the 3D dGPE, for a highly-elongated system ($\gamma=80$) and a moderately-elongated system ($\gamma=10$). In both cases, there is close agreement between the 3D dGPE and the variable-width approach, while the fixed-width approach has a noticeable deviation. The lower panels of Fig.~\ref{fig:result1} show the variational width $\sigma(z)$ according to Eq.~\eqref{eq:1dalg2}. For both cases, the departure of $\sigma$ from the harmonic oscillator length $\ell_\perp=\ell_z/\sqrt{\gamma}$ is significant, varying by up to 10-20\%.


\subsection{Arbitrary Polarisation}

If the dipole moments are polarised away from the $z$-axis ($\alpha\neq 0)$, the 1D DDI potential is not be analytic and must be computed numerically. Furthermore, the anisotropic nature of the DDIs will cause the cylindrical symmetry of the dipolar BEC to be broken. To accommodate this, we extend the ansatz for the wavefunction to include an anisotropic Gaussian profile in the $xy$ plane
\begin{equation}
\psi(\bm x,t)=\frac{f(z,t)}{\sqrt{\pi\sigma_x\sigma_y}}\exp\left\{-\frac{x^2}{2\sigma_x^2}-\frac{y^2}{2\sigma_y^2}\right\},
\end{equation}
where $\sigma_x(z,t)$ and $\sigma_y(z,t)$ are the variational widths.  Using this wavefunction, we evaluate the 3D DDI potential in Eq.~\eqref{eq:ddipotentialFourier},  the Fourier transforms in $x$ and $y$, and the inverse Fourier transforms in $k_x$ and $k_y$ direction. Only one of these integrals can be completed analytically. As such, the final equations will involve an inverse Fourier transform in either the $k_x$ or $k_y$ coordinate in Fourier space, and this will need to evaluated numerically.

Following the above approach, we integrate out the $x$ and $y$ directions in the action functional, and employ the Euler-Lagrange equations. This leads to three coupled equations, an intrego-differential equation for $\tilde{f}$ and two integral equations for $\sigma_x$ and $\sigma_y$.  In dimensionless form, these are,
\begin{widetext}
\begin{align}
i\partial_{\tilde{t}}\tilde{f}&=\left\{-\frac{1}{2}\partial_{\tilde{z}}^2+\frac{\tilde{z}^2}{2}+\frac{1}{4}\left(\frac{1}{\tilde{\sigma}_x^2}+\frac{1}{\tilde{\sigma}_y^2}\right)+\frac{\gamma^2}{4}(\tilde{\sigma}_x^2+\tilde{\sigma}_y^2)+\frac{\tilde{\beta}_{1\text{D}}|\tilde{f}|^2}{\tilde{\sigma}_x\tilde{\sigma_y}}\right.\notag\\
& \hspace{90px}\left. +\frac{3\sqrt{2\pi}\tilde{\beta}_{1\text{D}}\tilde{\epsilon}_{\text{dd}}}{\tilde{\sigma}_x}\mathcal{F}_{\tilde{z}}^{-1}\left[\int\text{d}\tilde{k}_y\,e^{-2\pi^2\tilde{k}_y^2\tilde{\sigma}_y^2}\left(\tilde{A}(q)\sin^2\alpha+\tilde{B}(q)\cos^2\alpha-\frac{1}{3}\right)\mathcal{F}_{\tilde{z}}\{|\tilde{f}|^2\}\right]\right\}\tilde{f},\label{eq:1dapp1}\\
0&=1-\gamma^2\tilde{\sigma}_x^4+\frac{\tilde{\beta}_{1\text{D}}|\tilde{f}|^2\tilde{\sigma}_x}{\tilde{\sigma}_y}\notag\\
& \hspace{94px}-3\sqrt{2\pi}\tilde{\beta}_{1\text{D}}\epsilon_{\text{dd}}\tilde{\sigma}_x\mathcal{F}_{\tilde{z}}^{-1}\left[\int\text{d}\tilde{k}_y\,e^{-2\pi^2\tilde{k}_y^2\tilde{\sigma}_y^2}\left(\tilde{C}(q)\sin^2\alpha+\tilde{D}(q)\cos^2\alpha+\frac{1}{3}\right)\mathcal{F}_{\tilde{z}}\{|\tilde{f}|^2\}\right],\label{eq:1dapp2}\\
0&=1-\gamma^2\tilde{\sigma}_y^4+\frac{\tilde{\beta}_{1\text{D}}|\tilde{f}|^2\tilde{\sigma}_y}{\tilde{\sigma}_x}\notag\\
& \hspace{75px}+\frac{12\sqrt{2}\pi^{5/2}\tilde{\sigma}_y^4\tilde{\beta}_{1\text{D}}\epsilon_{\text{dd}}}{\tilde{\sigma}_x}\mathcal{F}_{\tilde{z}}^{-1}\left[\int\text{d}\tilde{k}_y\,e^{-2\pi^2\tilde{k}_y^2\tilde{\sigma}_y^2}\left(\tilde{A}(q)\sin^2\alpha+\tilde{B}(q)\cos^2\alpha-\frac{1}{3}\right)\mathcal{F}_{\tilde{z}}\{|\tilde{f}|^2\}\right],\label{eq:1dapp3}
\end{align}
\end{widetext}
where 
\begin{align}
\tilde{A}(q)&=1-\sqrt{\pi}e^{q^2}q\,\text{erfc}(q),\\
\tilde{B}(q)&=\sqrt{2\pi^3}\tilde{k}_z^2\tilde{\sigma}_xe^{q^2}\text{erfc}(q)/\tilde{k}\,\\
\tilde{C}(q)&=2q^2-1-2\sqrt{\pi}q^3e^{q^2}\text{erfc}(q),\\
\tilde{D}(q)&=4\pi^2\tilde{k}_z^2\tilde{\sigma}_x^2(-1+\sqrt{\pi}qe^{q^2}\text{erfc}(q)),
\end{align}
and $\tilde{k}=\sqrt{\tilde{k}_y^2+\tilde{k}_z^2}$ and $q=\sqrt{2}\pi\sqrt{\tilde{k}_y^2+\tilde{k}_z^2}\,\tilde{\sigma}_x$, and $\text{erfc}(\cdot)$ is the complimentary error function. These are the variable-width GPE equations for the cigar-shaped dipolar BEC with arbitrary polarization angle $\alpha$.  For $\alpha=0$, for which the cylindrical symmetry of the BEC is preserved, the integrals can be completed analytically and the equations reduce to our previous $\alpha=0$ result, Eqs.~\eqref{eq:1ddiff2} and \eqref{eq:1dalg2}.

%

\begin{figure}[h!]
	\includegraphics[width=\linewidth]{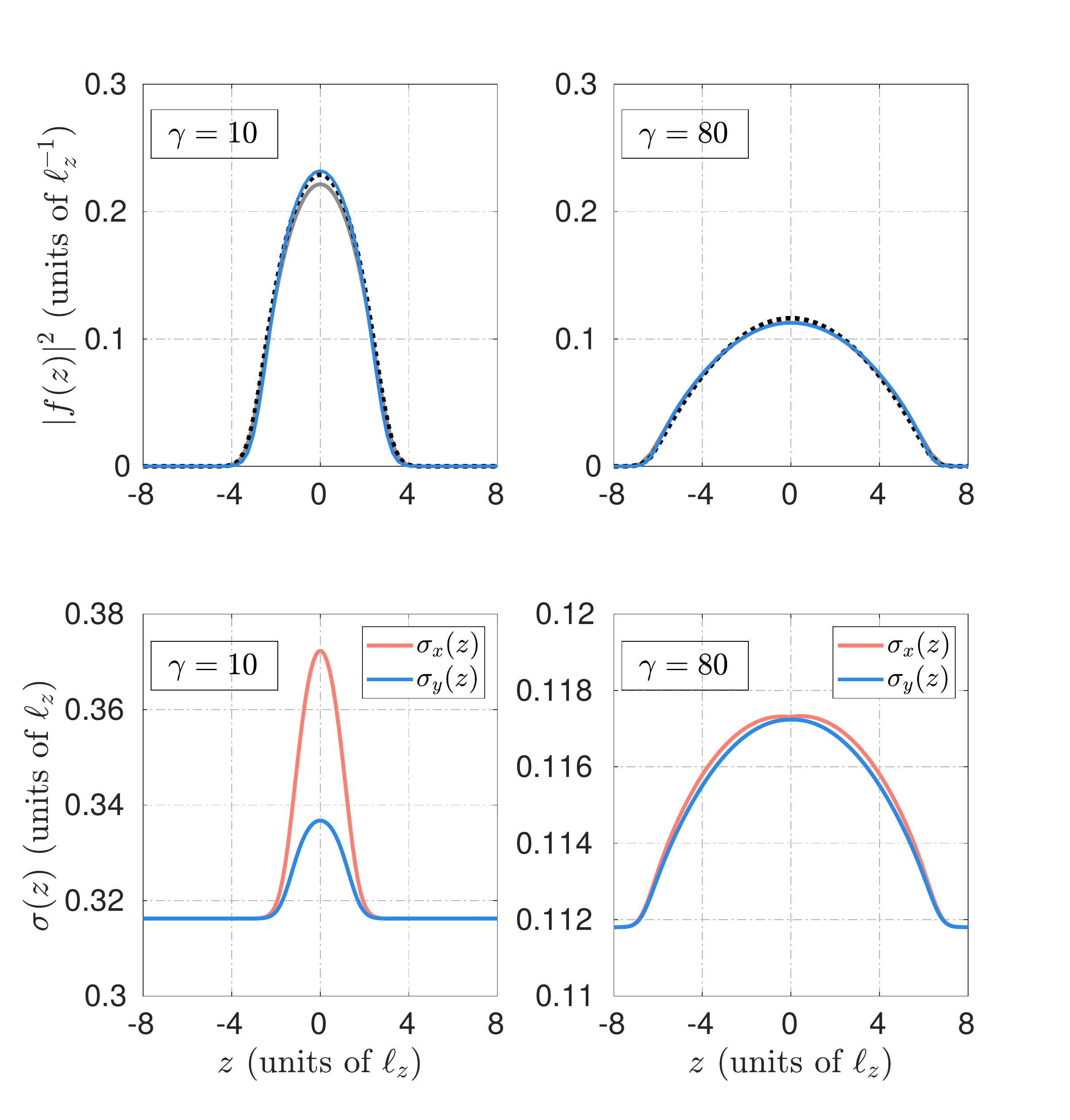}
	\caption{Cigar-shaped dipolar BECs with dipoles polarised at $\alpha=\pi/2$.  Top panels: Ground state density according to the variable-width approach (dotted black line), the fixed-width approach (solid grey line) and 3D GPE (solid blue line). Bottom panels: Variational widths $\sigma_x(z)$ and $\sigma_y(z)$ according to Eq.~\eqref{eq:1dapp2} and Eq.~\eqref{eq:1dapp3}. In all cases, $\epsilon_{\text{dd}}=0.9$ and $\tilde{\beta}_{1\text{D}}=2$.}
	\label{fig:result2}
\end{figure}

We proceed with $\alpha=\pi/2$, that is, dipoles polarized perpendicular to the primary axis of the trap, so as to break the symmetry about this axis.  In the top panels of Fig.~\ref{fig:result2} we compare the ground state density according to the variable-width approach, fixed-width approach and the 3D dGPE, for $\gamma=10$ and $80$.  Again, the variable-width approach more closely agrees with the 3D dGPE than the fixed-width approach.  For both values of $\gamma$, the widths again deviate significantly from $\ell_\perp$ (bottom panels), emphasizing that the fixed-width assumption is invalid.  
Furthermore, for $\gamma=10$ (bottom-left panel) we see strong anisotropy in the $xy$ plane, due to the dipole-induced magnetostriction of the condensate in the $x$ direction.  This emphasizes the importance of allowing the anisotropy in the transverse profile of the BEC. In the case for $\gamma=80$ (bottom-right panel), for which the system is more deeply in the 1D regime and the transverse trapping dominates over the interactions, the BEC approaches cylindrical symmetry.

These results are particularly useful, since the variable-width approach matches very closely with the 3D dGPE in the moderate trapping regime, $\gamma\sim 10$, where the transverse trap isn't extreme enough to diminish the effects of the DDIs. This excellent agreement only occurs because we take into account the breaking of the cylindrical symmetry. Without the variational widths, the approach would only work in the case where the effect of the DDIs are effectively nullified by the extreme trapping ratio $\gamma$, as seen in \cite{PhysRevA.82.043623}.

\begin{figure}[h]
	\includegraphics[width=0.9\linewidth]{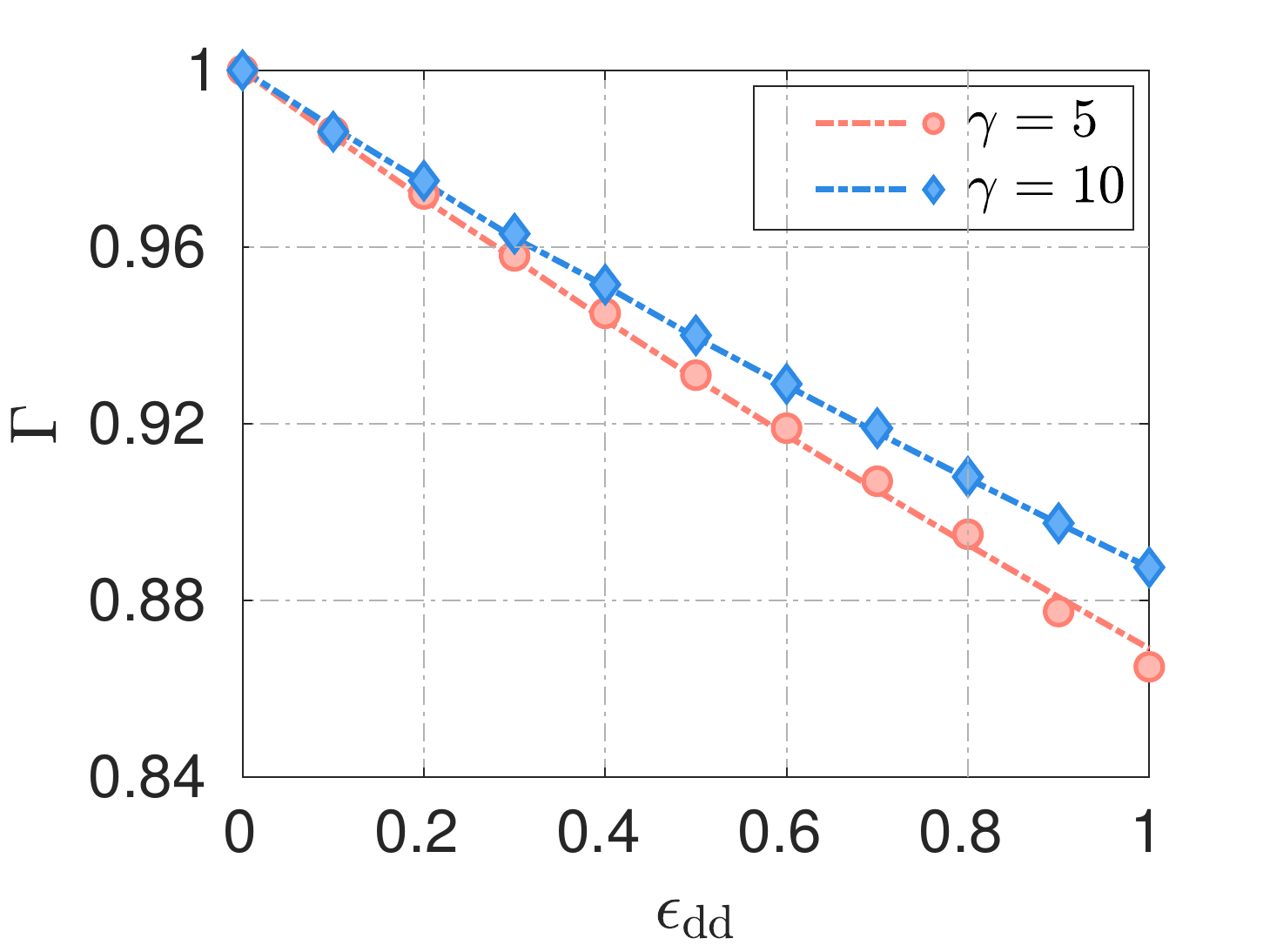}
	\caption{Transverse aspect ratio $\Gamma$ of the cigar-shaped dipolar BEC with dipoles polarized perpendicular to the long trap axis ($\alpha=\pi/2$), as a function of the DDI strength, $\epsilon_{\text{dd}}$. The markers show the predictions of the variable-width 1D model and the dashed lines represent the 3D dGPE solutions. We take $\tilde{\beta}_{1\text{D}}=2$.}
	\label{fig:resultaspect1}
\end{figure}

To characterise the anisotropy of the BEC in the $xy$ plane, we evaluate the transverse aspect ratio at $z=0$ (where $\sigma_x$ and $\sigma_y$ are maximal),
\begin{equation}
\Gamma=\frac{\max\{\tilde{\sigma}_y\}}{\max\{\tilde{\sigma}_x\}}. \label{eq:1daspect}
\end{equation}
In Fig.~\ref{fig:resultaspect1} we plot $\Gamma$ as a function of $\epsilon_{\rm dd}$ for two trap ratios, $\gamma=5$ and $10$.  We see the $\Gamma$ decreases monotonically in both cases, corresponding to the increasing magnetostriction along $x$ as the DDI strength $\epsilon_{\rm dd}$ is increased (relative to the contact interactions).  The transverse aspect ratio is closer to unity for the higher trap ratio $\gamma=10$.  This is because for large $\gamma$, the transverse energy effectively diminishes the interaction energy, such that transverse wavefunction becomes closer to the isotropic harmonic oscillator state. Remarkably, we see excellent agreement between the variable-width predictions (markers) and the 3D dGPE (dashed lines) throughout.  This demonstrates a marked improvement over the approach with isotropic fixed-width, for which $\Gamma$ would be fixed to unity throughout. 

%
%
\section{Variable-Width Two-Dimensional Equations}

We now consider a flattened dipolar BEC confined in an oblate harmonic potential with $\gamma \ll 1$.  Following Salasnich {\it et al.} \cite{PhysRevA.65.043614} we write the wavefunction in the following separable form,
\begin{align}
\psi(\bm x,t)&=f(x,y,t)\varphi(z,t;\eta(x,y,t))\notag\\
&=f(x,y,t)\frac{\exp\left\{-\displaystyle\frac{z^2}{2\eta(x,y,t)^2}\right\}}{\pi^{1/4}\sqrt{\eta(x,y,t)}},
\end{align}
where $f(x,y,t)$ is the two-dimensional wavefunction, and $\eta(x,y,t)$ is the variational width, which describes the extent of the BEC in the confined $z$-direction.  In the standard 2D reduction \cite{PhysRevA.82.043623}, $\eta$ is fixed to the oscillator length in the $z$-direction, $\ell_z=\sqrt{\hbar/(m\omega_z)}$.

To derive the variable-width 2D equations, we follow the same process as for the 1D case, except we now integrate out the $z$-component of the 3D action functional, and the $z$ and $k_z$ components of the Fourier and inverse Fourier transforms, respectively, in the DDI potential given by Eq.~\eqref{eq:ddiinteractionFourier}. Fortunately, the integrals can be completed analytically, so there is no particular need to separate out the $\alpha=0$ equations as a special case. Also, we assume that the axial wavefunction, $\varphi(z,t;\eta(x,y,t))$ is slowly varying in the $xy$-plane, allowing us to write $\nabla^2\varphi(z,t)\approx \partial_z^2\varphi(z,t)$. 

After the integration, the 2D action functional is, 
\begin{multline}
\mathcal{S}_{2\text{D}}=\int\text{d}x\,\text{d}y\,\text{d}t\,f^*\left\{i\hbar\partial_t+\frac{\hbar^2}{2m}\nabla_{\perp}^2-\frac{\hbar^2}{4m\eta^2}\frac{1}{4}m\omega_z^2\eta^2\right.\\-\left.\frac{1}{2}m\omega_{\perp}^2(x^2+y^2)-\frac{gN|f|^2}{2\sqrt{2\pi}\eta}-\frac{1}{2}\Phi_{\text{dd}}^{2\text{D}}\right\}f,\label{eq:2dactionfunctional}
\end{multline}
where $\nabla_{\perp}^2=\partial_x^2+\partial_y^2$.  $\Phi_{\text{dd}}^{2\text{D}}$ is the 2D DDI term, derived from Eq.~\eqref{eq:ddiinteractionFourier} by integrating out the $z$ and $k_z$-components of the Fourier transforms, and has the form,
\begin{multline}
\Phi_{\text{dd}}^{2\text{D}}=\frac{C_{\text{dd}}N}{\sqrt{2\pi}\eta}\mathcal{F}_{x,y}^{-1}\Bigg\{\Big[\sin^2\alpha\, G(\xi)+\cos^2\alpha \,F(\xi)\\
-\frac{1}{3}\Big]\mathcal{F}_{x,y}\left\{|f|^2\right\}\Bigg\}f,\label{eq:2dddipotential}
\end{multline}
where $\mathcal{F}_{x,y}$ ($\mathcal{F}^{-1}_{x,y}$) is the 2D Fourier (inverse Fourier) transform, and 
\begin{align*}
G(\xi)&=\sqrt{2}\pi^{3/2}k_x^2\eta e^{\xi^2}\,\text{erfc}(\xi)/k_{\perp}\\
F(\xi)&=1-\sqrt{\pi}\xi e^{\xi}\,\text{erfc}(\xi),\\
\end{align*}
where $\xi=\sqrt{2}\pi\eta k_{\perp}$, $k_{\perp}=\sqrt{k_x^2+k_y^2}$, and $\text{erfc}(\cdot)$ is the complimentary error function. We can see that this 2D DDI potential has a clear dependence on the polarisation direction $\alpha$.

An important aspect of this DDI potential is the asymmetric dependence on the Fourier space coordinates $k_x$ and $k_y$, which is in contrast to the 1D DDI potential, which had a dependence only on $k_z$ (after integration) and hence maintained symmetry. This asymmetry takes into effect magnetostriction already, and as such, a fixed-width approach to calculating the ground state for dipoles polarised away from the axis of symmetry will prove to be quite accurate as compared to the 3D dGPE. Furthermore, a calculation of the transverse aspect ratio can be reached without the inclusion of a variational width. The inclusion of $\eta(x,y,t)$, which we recall is a measure of the width in the $z$-direction, simply increases the accuracy of these predictions. This was not the case for the 1D BEC, as the fixed-width approach set the aspect ratio to unity, and no magnetrostriction was observed unless we let the orthogonal widths $\tilde{\sigma}_x$ and $\tilde{\sigma}_y$ vary.

For this 2D case, we employ units defined by the trapping in the transverse plane, with a time-scale $1/\omega_{\perp}$ and a length scale $\ell_{\perp}=\sqrt{\hbar/(m\omega_{\perp})}$, and transform our dimensions to the dimensionless forms according to,
\begin{align}
\begin{split}
&\tilde{t}=t\omega_{\perp}, \ (\tilde{x},\tilde{y})=\frac{(x,y)}{\ell_{\perp}}, \ \tilde{a}_s=\frac{a_s}{\ell_{\perp}}\\
&\tilde{\eta}=\frac{\eta}{\ell_{\perp}}, \ (\tilde{k}_x,\tilde{k}_y)=\frac{(k_x,k_y)}{\ell_{\perp}}.
\end{split}
\end{align}
The dimensionless 2D wavefunction becomes $\tilde{f}=\ell_{\perp}f$.  The Euler-Lagrange equations with respect to the 2D wavefunction and variable width $\tilde{\eta}$ giving the following equations,
\begin{widetext}
\begin{align}
i\partial_{\tilde{t}}\tilde{f}&=\left\{-\frac{1}{2}\tilde{\nabla}_{\perp}^2+\frac{1}{4\tilde{\eta}^2}+\frac{1}{2}(\tilde{x}^2+\tilde{y}^2)+\frac{\tilde{\eta}^2}{4\gamma^2}+\tilde{\beta}_{2\text{D}}\frac{|\tilde{f}|^2}{\tilde{\eta}}\right.\notag\\
& \left.\hspace{190px}+\frac{3\epsilon_{\text{dd}}\tilde{\beta}_{2\text{D}}}{2\tilde{\eta}}\mathcal{F}_{\tilde{x},\tilde{y}}^{-1}\left[\left(\tilde{F}_{\perp}\cos^2\alpha+\tilde{F}_{\parallel}\sin^2\alpha-\frac{1}{3}\right)\mathcal{F}_{\tilde{x},\tilde{y}}\{|\tilde{f}|^2\}\right]\right\}\tilde{f}, \label{eq:2ddiff1}\\
0&=1-3\tilde{\beta}_{2\text{D}}\epsilon_{\text{dd}}\tilde{\eta}\mathcal{F}^{-1}\left[\left(4\pi^2\tilde{k}_x^2\tilde{\eta}^2\tilde{G}_{\parallel}\sin^2\alpha+\tilde{G}_{\perp}\cos^2\alpha+\frac{1}{3}\right)\mathcal{F}_{\tilde{x},\tilde{y}}\{|\tilde{f}|^2\}\right]-\frac{\tilde{\eta}^4}{\gamma^2}+\tilde{\beta}_{2\text{D}}|\tilde{f}|^2\tilde{\eta},\label{eq:2dalg1}
\end{align}
\end{widetext}
where $\tilde{\beta}_{2\text{D}}=2\sqrt{2\pi}\tilde{a}_sN$, and 
\begin{align}
\tilde{F}_{\perp}&=1-\sqrt{\pi}\xi e^{\xi^2}\text{erfc}(\xi),\label{eq:2dterm1}\\
\tilde{F}_{\parallel}&=\sqrt{2}\pi^{3/2}\tilde{k}_x^2\tilde{\eta}e^{\xi^2}\text{erfc}(\xi)/\tilde{k}_{\perp},\label{eq:2dterm2}\\
\tilde{G}_{\perp}&=2\xi^2-1-2\sqrt{\pi}\xi^3e^{\xi^2}\text{erfc}(\xi),\label{eq:2dterm3}\\
\tilde{G}_{\parallel}&=\sqrt{\pi}\xi e^{\xi^2}\text{erfc}(\xi)-1.\label{eq:2dterm4}
\end{align}
Here $\mathcal{F}_{\tilde{x},\tilde{y}}$ ($\mathcal{F}_{\tilde{x},\tilde{y}}^{-1}$) denotes the Fourier (inverse Fourier) in the $\tilde{x}$ an $\tilde{y}$ ($\tilde{k}_x$ and $\tilde{k}_y$) directions. These equations have a similar form to the 1D equations for the arbitrary polarization direction (Eqs.~\eqref{eq:1dapp1}, ~\eqref{eq:1dapp2}, and \eqref{eq:1dapp3}), maintaining a direct dependence upon $\alpha$. There is only one algebraic equation since $\tilde{\eta}$ is a measure of the physical extent of the BEC in the $z$-direction only. 
\begin{figure}[h]
	\includegraphics[width=\linewidth]{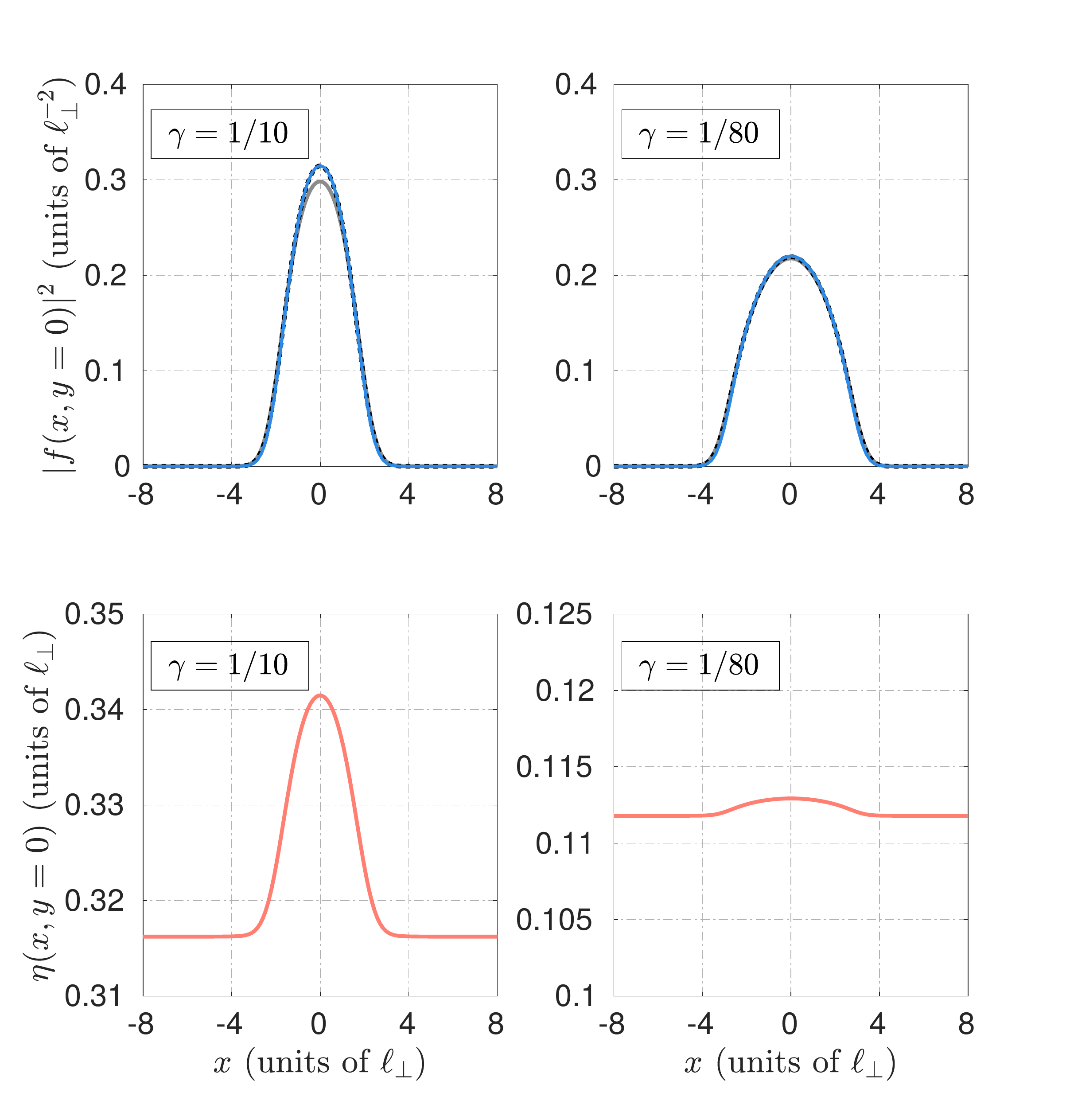}
	\caption{Pancake-shaped BEC with dipoles polarised along $z$.  Top panels: Ground state density profiles from the variable-width approach (black dotted line), the fixed-width approach (solid grey line), and the 3D dGPE (blue solid line).  Bottom panels:  Profile of the variational width $\tilde{\eta}$ along the $x$ axis. We take $\tilde{g}_{2\text{D}}=2\sqrt{2\pi}$, $\epsilon_{\text{dd}}=0.9$ and $\alpha=0$.}  
	\label{fig:result4}
\end{figure}

First we consider the dipoles to be polarized in $z$, $\alpha=0$.
In Fig.~\ref{fig:result4} (top panels) we plot the ground state density profile along the $x$-axis, $|\tilde{f}(\tilde{x},\tilde{y}=0)|^2$, for trap ratios $\gamma=1/10$ and $1/80$.  The variable-width results (black dotted line) are in excellent agreement with those of the 3D dGPE (blue solid line) for both trap ratios.  For $\gamma=1/10$, the fixed-width approach noticeably underestimates the correct density. This discrepany is reduced for $\gamma=1/80$, as the axial trapping energy dominates over the interaction energy of the BEC. 

\begin{figure}[h]
\includegraphics[width=\linewidth]{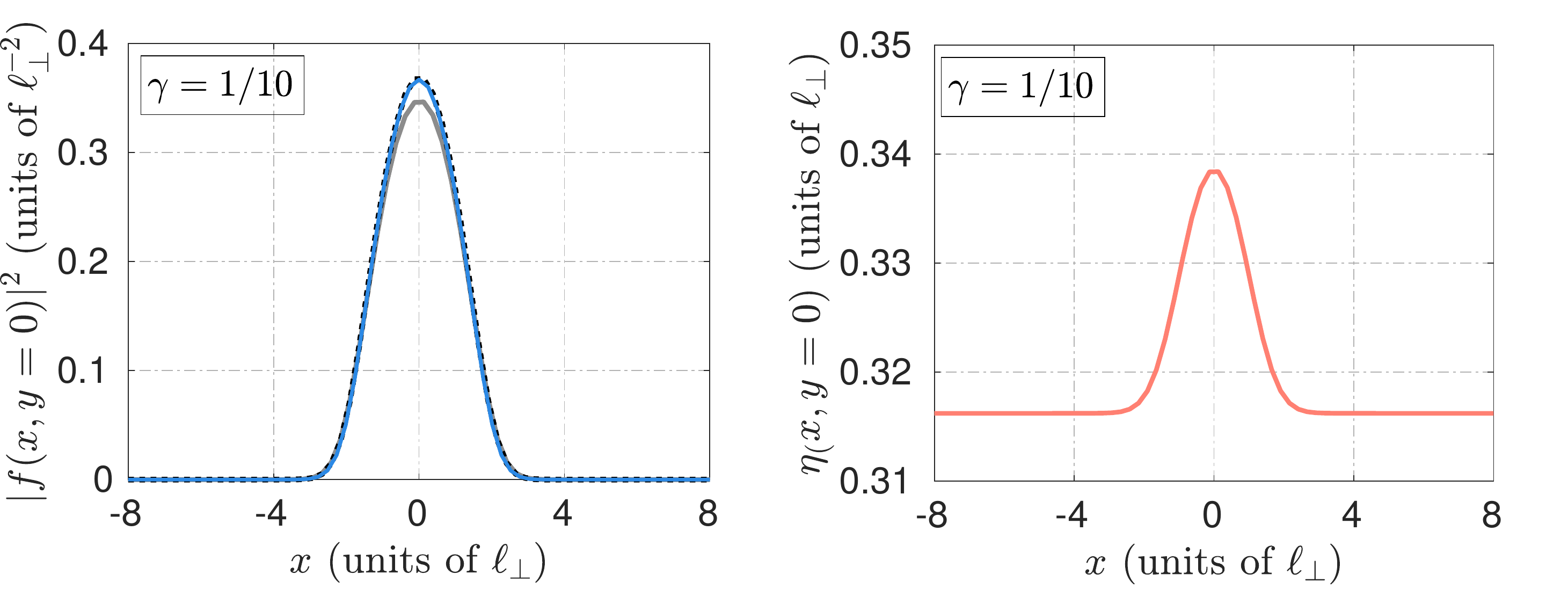}
\caption{Pancake-shaped BEC with dipoles polarised along $x$.  Left: Ground state density profile along the $x$-axis according to the variable-width approach (dotted black line), fixed-width approach (solid grey line) and 3D dGPE (solid blue line).  Right:  Profile of the variational width along the $x$-axis, $\eta(x,y=0)$.  We take $\tilde{\beta}_{2\text{D}}=2\sqrt{2\pi}$, $\epsilon_{\text{dd}}=0.9$ and $\gamma=1/10$.}
\label{fig:result5}
\end{figure}

\begin{figure}
	\includegraphics[width=0.9\linewidth]{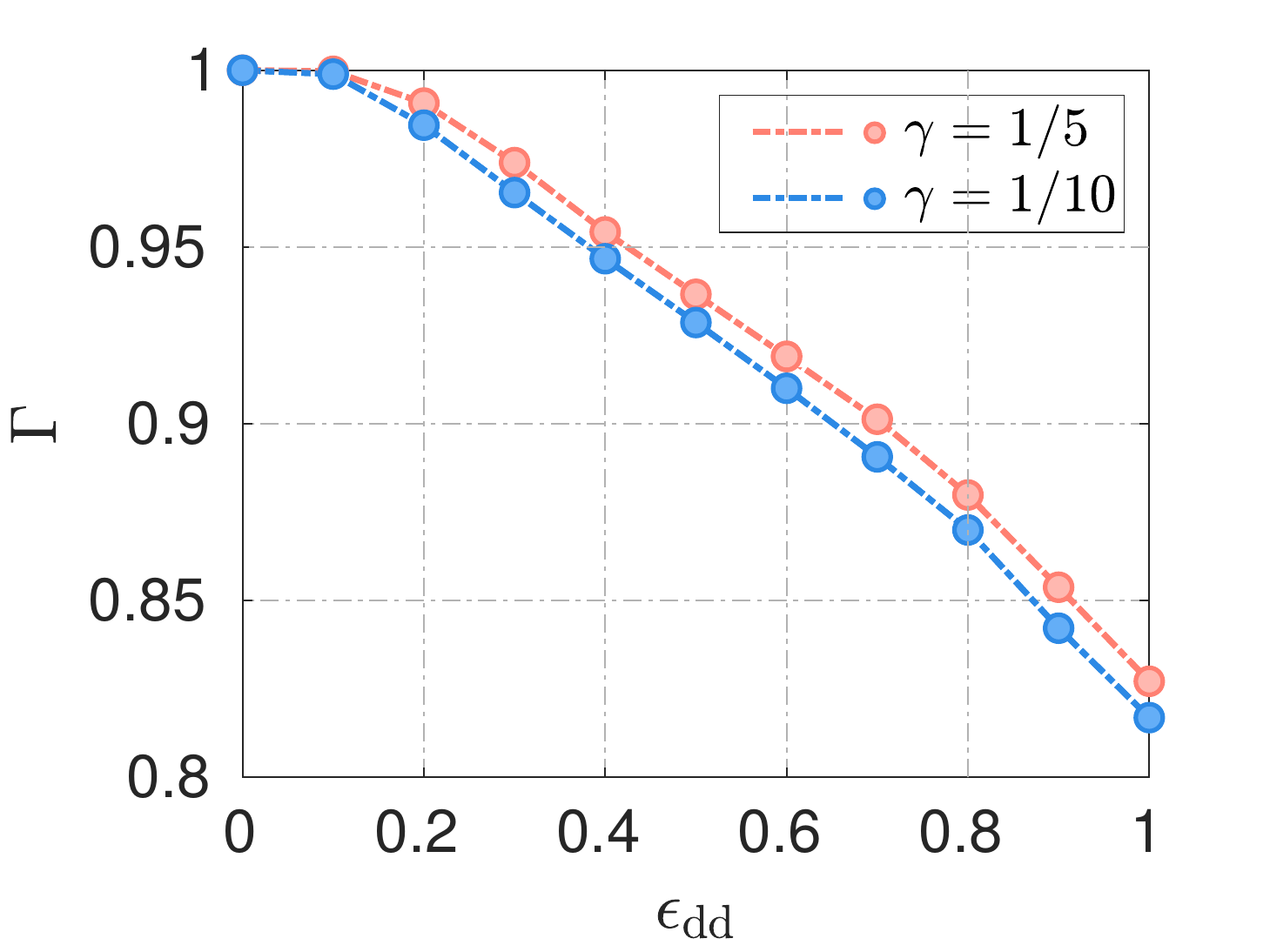}
	\caption{Transverse aspect ratio of a pancake-shaped dipolar BEC as a function of the DDI strength, $\epsilon_{\text{dd}}$, with the dipoles polarized along $x$, according to the variable-width approach (markers) and 3D dGPE (dashed lines).  }
	\label{fig:result6}
\end{figure}

We next consider the dipoles to be polarized along $x$, $\alpha=\pi/2$, with the results shown in Fig.~\ref{fig:result5}.  Again, the variable width approach accurately reproduces the results of the 3D dGPE, while there is a noticeable deviation with the fixed-width approach.  
Finally, we consider the transverse aspect ratio of the BEC, $\Gamma$, measured from the physical extent of the wavefunction, $\tilde{f}(\tilde{x},\tilde{y})$ in the $xy$ plane.  The variation of $\Gamma$ with $\epsilon_{\rm dd}$ is shown in Fig.~\ref{fig:result6} for two trap ratios.  The aspect ratio decreases as the dipoles are increased in strength, due to the increasing magnetostriction  in the $x$ direction.  The variable-width predictions closely match 3D dGPE throughout. 

%

\vspace{10px}

\section{Conclusions}

In this paper, we have derived effective 1D and 2D mean-field equations describing cigar-shaped and pancake-shaped dipolar Bose-Einstein condensates with arbitrary polarization angle.  Previous works in this direction have obtained effective 1D and 2D equations by assuming that the condensate profile is fixed to the ground Gaussian harmonic oscillator state in the tightly-confined direction; this ``fixed-width" approach is valid for strictly quasi-1D and quasi-2D BECs.  Here we extend on this by allowing the Gaussian profile to have a variational width which is solved self-consistently; this ``variable-width" approach leads to effective 1D and 2D equations for the low-dimensional wavefunction, coupled to equations for the widths of the tightly-confined directions.

We compute the ground state of cigar-shaped and pancake-shaped dipolar BECs, and show that our effective equations give excellent agreement with the full 3D dGPE solutions over a range of trap ratios, polarization angles and interaction strengths.  In particular, they provide a significant improvement over the fixed-width approach in that they remain accurate and valid as the system geometry is relaxed away from the strict quasi-1D and quasi-2D regimes.  In addition to this, they are more efficient to numerically solve than the 3D dGPE. 

%

These approaches may provide a convenient means to model the high-dimensional physics of the roton and dipolar supersolid phases \cite{PhysRevA.75.053604,PhysRevLett.100.245302,PhysRevA.82.023622,PhysRevLett.122.130405,PhysRevX.9.011051,PhysRevX.9.021012} within a low-dimensional model; in order to study the supersolid phases, our approach would need to be extended to incorporate the role of quantum fluctuations.  A further interesting case would be to use this approach to study the dynamics of dark solitons in cigar-shaped BECs.  A significant discrepancy has been noted between the prediction of the 3D GPE and standard 1D reduction \cite{PhysRevA.95.063622}, and the success of the NPSE in capturing observed dark soliton oscillations in non-dipolar BECs \cite{PhysRevLett.101.130401} suggests that this approach could well reconcile this discrepancy.

\section*{Acknowledgments}

M. J. K. is supported by an Australian Government Research Training Program (RTP) Scholarship and A. M. M. would like to thank the Australian Research Council (Grant No. LE180100142) for support. T. B. and N. G. P. thank the Engineering and Physical Sciences Research Council (Grant No. EP/M005127/1) for support.

\bibliography{low_dim_eqns.bib}

\end{document}